\documentclass[sigplan,nonacm]{acmart}
\usepackage{amsmath,amsfonts}
\usepackage{physics}
\usepackage{algorithmic}
\usepackage{algorithm}
\usepackage{graphicx}
\usepackage{subcaption}
\usepackage{textcomp}
\usepackage{xcolor}
\usepackage{diagbox}
\usepackage{multirow}
\usepackage{multicol}
\usepackage{makecell}
\usepackage{enumitem}
\usepackage{hyperref}
\hypersetup{}

\newtheorem{theorem}{Theorem}

\newcommand{\jz}[1]{\textcolor{purple}{[JZ: #1]}}
\newcommand{\myCompilerName}{QTurbo} 
\newcommand{\myCompilerNameSpace}{QTurbo }

\begin{document}
\title{\myCompilerName: A Robust and Efficient Compiler for Analog Quantum Simulation}

\author{Junyu Zhou}
\affiliation{%
  \institution{University of Pennsylvania}
  \city{Philadelphia}
  \country{USA}
}
\email{junyuzh@seas.upenn.edu}

\author{Yuhao Liu}
\affiliation{%
  \institution{University of Pennsylvania}
  \city{Philadelphia}
  \country{USA}
}
\email{liuyuhao@seas.upenn.edu}

\author{Shize Che}
\affiliation{%
  \institution{University of Pennsylvania}
  \city{Philadelphia}
  \country{USA}
}
\email{shizeche@seas.upenn.edu}

\author{Anupam Mitra}
\affiliation{%
  \institution{Lawrence Berkeley National Laboratory}
  \city{Berkeley}
  \country{USA}
}
\email{AnupamMitra@lbl.gov}

\author{Efekan K\"okc\"u}
\affiliation{%
  \institution{Lawrence Berkeley National Laboratory}
  \city{Berkeley}
  \country{USA}
}
\email{ekokcu@lbl.gov}

\author{Ermal Rrapaj}
\affiliation{%
  \institution{Lawrence Berkeley National Laboratory}
  \city{Berkeley}
  \country{USA}
}
\email{ermalrrapaj@lbl.gov}

\author{Costin Iancu}
\affiliation{%
  \institution{Lawrence Berkeley National Laboratory}
  \city{Berkeley}
  \country{USA}
}
\email{cciancu@lbl.gov}

\author{Gushu Li}
\affiliation{%
  \institution{University of Pennsylvania}
  \city{Philadelphia}
  \country{USA}
}
\email{gushuli@seas.upenn.edu}

\begin{abstract}
Analog quantum simulation leverages native hardware dynamics to emulate complex quantum systems with great efficiency by bypassing the quantum circuit abstraction. However, conventional compilation methods for analog simulators are typically labor-intensive, prone to errors, and computationally demanding. This paper introduces \myCompilerName, a powerful analog quantum simulation compiler designed to significantly enhance compilation efficiency and optimize hardware execution time. By generating precise and noise-resilient pulse schedules, our approach ensures greater accuracy and reliability, outperforming the existing state-of-the-art approach.

\end{abstract}

\maketitle % should come after the abstract

% add the paper content here
\section{Introduction} \label{Sec:Introduction}

Quantum computing leverages quantum mechanics to solve complex problems beyond classical capabilities, with applications in quantum chemistry~\cite{mcquarrie2008quantum}, combinatorial optimization~\cite{farhi2014qaoa}, and cryptography~\cite{shor1994algorithm,shor1997polynomial}. Among these, quantum Hamiltonian simulation~\cite{feynman1982simulating} is one of the most fundamental and promising problems, crucial for understanding quantum many-body systems, modeling chemical reactions, and designing new materials~\cite{mcquarrie2008quantum,altland2006condensed,barger2012physics,Patwardhan:2020,Cirigliano:2024}. There are two main approaches: \textit{digital} and \textit{analog}. Digital quantum simulation~\cite{fauseweh2024quantum} uses quantum gates to construct a circuit that approximates Hamiltonian time evolution, often via Suzuki-Trotter decomposition~\cite{hatano2005suzuki}. However, digital quantum simulation requires millions of gate operations even for medium-sized systems. For example, Childs et al.~\cite{childs2018toward} shows that simulating a simple system of around 100 qubits requires around $10^{10}$ logical gates before error correction encoding.

\begin{figure}[t]
    \centering
    \includegraphics[width=0.95\linewidth]{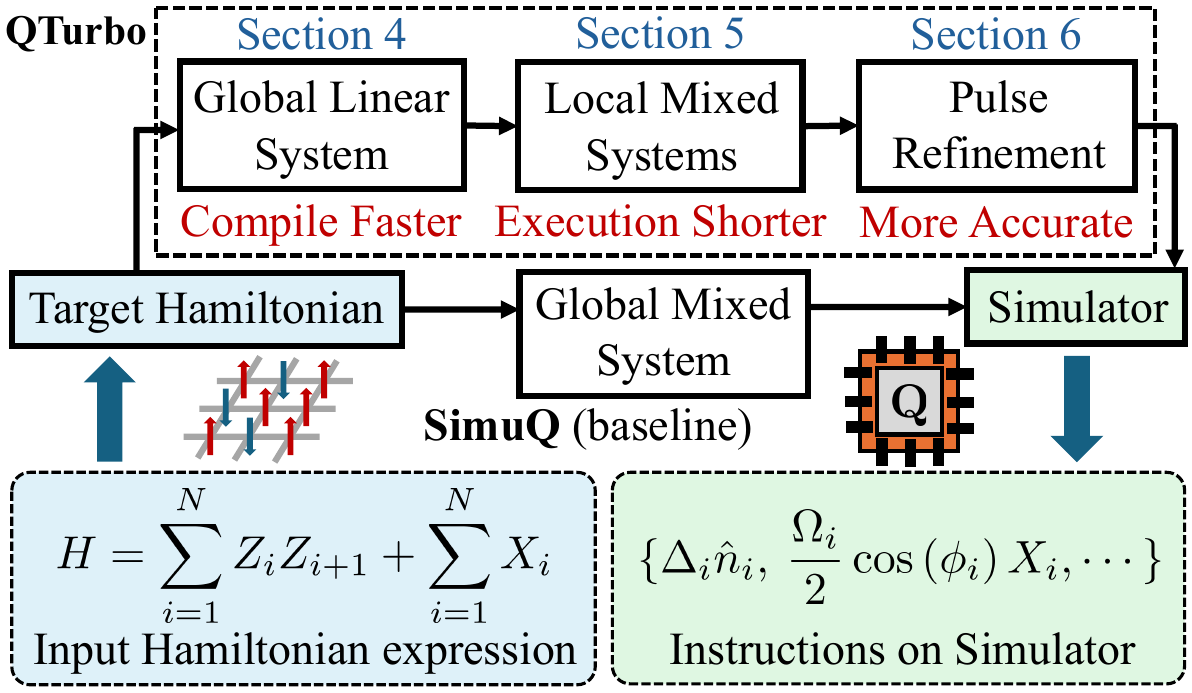}
    % \vspace{-5pt}
    \caption{Comparision between the proposed \myCompilerNameSpace compiler and the baseline SimuQ~\cite{peng2024simuq}}
    % \vspace{-5pt}
    \label{fig:AQS}
    \Description{}
\end{figure}

On the other hand, analog quantum simulation~\cite{arguello2019analogue} has garnered increasing interest due to its ability to emulate the behavior of a target quantum system directly. Unlike the digital approach, analog quantum simulation bypasses the quantum circuit abstraction and directly exploits the native tunable Hamiltonian of the quantum hardware~\cite{wurtz2023aquila} to mimic a specific Hamiltonian’s dynamics continuously. 
%This direct hardware-level mimicry allows for more scalable and noise-resilient simulations, as it avoids the compounding of gate errors inherent in digital schemes~\cite{hatano2005suzuki}. 
Analog quantum simulation has thus become a powerful tool for studying complex quantum phenomena. 
For example, QuEra has studied phases of matter~\cite{ebadi2021quantum} and solved combinatorial optimization problems~\cite{ebadi2022quantum} on its Rydberg atom-based 256-qubit computer Aquila~\cite{wurtz2023aquila}.
Recently, Google also studied physics experiments with analog quantum simulation on its superconducting quantum processor~\cite{andersen2025thermalization}.
%Recent advancements in programmable analog simulators—such as QuEra’s Rydberg atom-based 256-qubit computer, Aquila~\cite{wurtz2023aquila}—underscore the potential of analog approaches for practical quantum simulations. \jz{Experimental studies have recently demonstrated the efficacy of analog quantum simulation in probing physical systems~\cite{andersen2025thermalization, braumuller2017analog, ebadi2021quantum} and tackling optimization problems~\cite{ebadi2022quantum}. 
Additionally, error mitigation techniques have been investigated for analog quantum simulation~\cite{meher2024error}.

Despite the promising potential of analog quantum simulation, the development of compilation techniques for analog simulation programs remains in its early stages. Currently, the only publicly available software framework is SimuQ~\cite{peng2024simuq}, to the best of our knowledge. As illustrated in Fig. \ref{fig:AQS}, SimuQ maps a target quantum system to a quantum simulator's instruction set by formulating and solving a global mixed equation system, ultimately producing pulse schedules for the analog simulation device. This framework establishes a unified, solver-based compilation pipeline that translates user-defined Hamiltonians into low-level hardware controls. However, the SimuQ compiler faces two critical challenges that limit its efficiency and reliability. \textbf{First,} the equation system used by SimuQ to determine pulse schedules solves all variables simultaneously, inherently leading to a large-scale mixed continuous–binary optimization problem and an exponentially large search space with a very long compilation time. 
%However, some instructions on simulator are actually separable and do not influence each other. 
\textbf{Second,} the length of the compiled analog pulse sequence is not deterministic, usually far from optimal, and can vary unpredictably with different solver conditions. 
In some cases, SimuQ even fails to yield a solution.
%the solution results in an evolution time that is too long for NISQ devices, causing decoherence before the simulation can be completed. 

% Traditionally, programming an analog simulator meant manually calculating control parameter schedules (e.g. laser intensities, frequencies) and tuning them on the quantum hardware for each specific problem – a labor-intensive approach that lacks scalability and is prone to human error~\cite{????????}. 

In this paper, our objective is to develop a fast and effective compiler that can compile and optimize the mapping from the input Hamiltonian onto an analog quantum simulator.
Our key observation is that there exists a \textit{hierarchical structure} in the seemingly monolithic, large mixed equation system in the analog simulation compilation.
Instead of directly solving all the Hamiltonian controls, we can decompose the compilation into multiple stages with intermediate variables.
Moreover, the tunable terms on the analog quantum simulator are not always dependent on each other, and it is possible to turn the large equation system into multiple smaller \textit{local} equation systems that are easier to solve.

%\jz{The key insight is leveraging the separability within the global mixed equation system constructed by SimuQ. By incorporating simulator's constraints, we can decompose the equations into smaller, independent parts, allowing for the determination of an optimal pulse duration.}

To this end, we propose \myCompilerName, an efficient and robust compilation framework for analog quantum simulation. 
%In contrast to SimuQ, \myCompilerNameSpace can exponentially reduce the compilation time for a given analog quantum program while deterministically returning an optimal pulse duration. 
The workflow of \myCompilerNameSpace is illustrated in Fig. \ref{fig:AQS}. \textbf{First}, \myCompilerNameSpace optimizes the construction of the equation system for pulse design by decomposing the global mixed equation system into a \textit{global linear} equation system and several \textit{localized mixed} equation systems. This approach substantially enhances the equation-solving process by reducing computational complexity and improving scalability. \textbf{Second}, when solving the localized mixed equation systems, \myCompilerNameSpace enhances pulse execution efficiency by leveraging variable constraints—such as the maximum amplitude of a laser beam—to determine the shortest achievable overall evolution time, thereby ensuring operation within coherence constraints and maximizing simulation reliability. \textbf{Third}, \myCompilerNameSpace introduces an iterative refinement process to minimize error propagation across both linear and localized mixed equation systems, thereby reducing the error of compiled pulse schedules and improving fidelity in analog quantum simulations.
Consequently, \myCompilerNameSpace can efficiently handle compiling large-scale quantum system onto analog quantum simulator with shorter pulse and higher approximation accuracy. %, consistently generating pulses that are both reliable and robust against noise for actual device execution.

Our experimental results indicate that, compared with the baseline SimuQ~\cite{peng2024simuq}, \myCompilerNameSpace achieves around $600\times$ (up to $1600\times$) acceleration in compilation time and scales effectively to larger systems. Additionally, the generated pulses are $51\%$ (up to $90\%$) shorter, and our pulse schedules demonstrate a $72\%$ (up to $100\%$) improvement in compilation accuracy. These enhancements directly translate into significant noise suppression on real device. We test two target quantum systems using QuEra's Aquila analog quantum computer, achieving an average error reduction of $51\%$, with a maximum reduction of up to $94\%$  in the measurement results.

% Furthermore, the shorter pulses generated by \myCompilerNameSpace enable the execution of tasks that were previously infeasible due to extended evolution times.

Our major contributions can be summarized as follows:
\begin{enumerate}
    \item We proposed \myCompilerName, an efficient and effective compiler optimization framework for analog quantum simulation protocol.
    \item \myCompilerNameSpace introduces several key compiler designs and optimizations that can reduce compilation time, reduce pulse execution time, and improve compilation accuracy.
    \item Our evaluation demonstrates that \myCompilerNameSpace outperforms the baseline SimuQ by significantly reducing both compilation time and actual device execution time across various benchmarks and architectures, and markedly enhancing actual device performance.
\end{enumerate}
\section{Preliminary} \label{Sec:Preliminary}
This section introduces the background knowledge of analog quantum simulation and the state-of-the-art compilation framework. Interested readers are encouraged to read~\cite{georgescu2014quantum, peng2024simuq, nielsen2010quantum} for further detailed information. 

\subsection{Analog Quantum Simulation} \label{Sec:AnalogQuantumSimulation}
The evolution of a quantum system is governed by its Hamiltonian $H$ according to the Schr\"{o}dinger equation:
\begin{equation*}
    i\hbar \frac{d}{dt} \lvert \Psi(t) \rangle = H \lvert \Psi(t) \rangle.
\end{equation*}
Analog quantum simulation (AQS) is a strategy used to study the behavior of such complex quantum systems by constructing another controllable quantum system that directly mimics the behavior of the system of interest. 

Figure~\ref{fig:AQS} shows this simulation process. Suppose $H_{tar}$ is the Hamiltonian of the target physical system to be simulated. We need to tune the pulses in the quantum processor (the simulator) to adjust its Hamiltonian $H_{sim}$, and approximate the target Hamiltonian $H_{tar}$ using $H_{sim}$. Finally, the measurement result of this quantum device can be used to infer the information about target system.

% \begin{figure}[t]
%     \centering
%     \includegraphics[width=1\linewidth]{fig/AQS.pdf}
%     % \vspace{-5pt}
%     \caption{Analog Quantum Simulation}
%     % \vspace{-5pt}
%     \label{fig:AQS}
% \end{figure}

\textbf{Warm-Up Example} Here, we provide a concrete example. 
Suppose the target system has three qubits and its Hamiltonian is:
\begin{equation*}
    H_{tar} = Z_{1}Z_{2} + X_{2}X_{3},
\end{equation*}
and evolves for a period of time $T=1$. $Z_{1}Z_{2}$ and $X_{2}X_{3}$ are the Pauli strings~\cite{nielsen2010quantum}. The subscript represents the index of the qubit that the corresponding operator acts on. For qubits whose indices do not explicitly appear in the Pauli string, identity operators are implicitly applied. For example, $Z_{1}Z_{2}$ is actually $Z_1Z_2I_3$. In the rest of this paper, we always omit the identity operators for simplicity.

Suppose the simulator contains the following tunable terms:
\begin{equation*}
    \{a_{1} \cdot Z_1 Z_2, \enspace a_2 \cdot Z_2 Z_3, \enspace b_1 \cdot X_1 X_2, \enspace b_2 \cdot X_2 X_3\},
\end{equation*}
where $a_i$ and $b_i$ are the amplitudes of pulses that users can increase or decrease to control the interaction strength of Hamiltonian terms $Z_{i}Z_{j}$ and $X_{i}X_{j}$ in the set. The amplitudes $a_i$ and $b_i$ are also called variables of the analog quantum program. This set of instructions is also known as the Abstract Analog Instruction Set (AAIS)~\cite{peng2024simuq} of such a quantum simulator. By simply setting $a_1, \enspace b_2$ to 1, and $a_2, \enspace b_1$ to 0, one can get the Hamiltonian of simulator:
\begin{equation*}
    H_{sim} = Z_{1}Z_{2} + X_{2}X_{3},
\end{equation*}
which is exactly the same as $H_{tar}$. Letting the simulator evolving for $T=1$ finishes the simulation task.

%\textbf{AAIS of Real Analog Quantum Simulators}
The AAIS of a real analog quantum simulator can be much more complicated.
We introduce the two most commonly used AAISs.

\subsubsection{Rydberg AAIS} Rydberg AAIS is the instruction set abstraction from a multi-atoms Rydberg device (e.g., the QuEra's Aquila~\cite{wurtz2023aquila}):
\begin{equation*}
   \{ \frac{C_6}{|x_i - x_j|^6}\hat{n}_i \hat{n}_j, \hspace{0.2em} -\Delta_i\hat{n}_i, \hspace{0.2em} \frac{\Omega_i}{2} \cos\left(\phi_i\right) X_i, \hspace{0.2em} -\frac{\Omega_i}{2} \sin\left(\phi_i\right) Y_i\},
\end{equation*}
where $\hat{n}_i=(I-Z_{i})/2$, and $X_i$, $Y_{i}$, $Z_{i}$ are Pauli X, Y, Z operators on the $i$th atom for all $i$ and $j$. The first instruction is the Van der Waals interaction between atoms. $C_{6} = 862690$ $MHz$ $\mu$m$^6$ is a constant, $x_i$ and $x_j$ are the atom positions. One cannot change the atom positions if quantum program starts to execute, thus the $x_i$s are called \textbf{runtime fixed variables}. The second instruction is called detuning term with amplitude $\Delta_i$. The last two instructions are called Rabi drive with Rabi amplitude $\Omega_i$ and phase $\phi_i$. $\Delta_i$, $\Omega_i$ and $\phi_i$ can vary during the program execution, thus they are \textbf{runtime dynamic variables}. This AAIS can be realized on the QuEra's device Aquila~\cite{wurtz2023aquila}, currently they only allow a global control of $\Delta_i$, $\Omega_i$ and $\phi_i$. We denote runtime fixed variables and runtime dynamic variables together as \textbf{amplitude variables} since they directly control the amplitude of instruction Hamiltonian terms.

\subsubsection{Heisenberg AAIS} Heisenberg AAIS is the instruction set abstraction from superconducting or trapped ion device:
\begin{equation*}
   \{ a^{P_i} \cdot P_i, \hspace{0.2em} a^{P_iP_j} \cdot P_iP_j\},
\end{equation*}
where $P \in \{X, Y, Z\}$ is Pauli operator and $a^{P_i}$, $a^{P_iP_j}$ are amplitudes of those Pauli strings. $a^{P_iP_j}$ may depends on the connectivity of the quantum device. All the amplitudes here are runtime dynamic variables. 
Although IBM recently removed pulse-level access to all its cloud quantum processors~\cite{ibmpulse} and there is not yet pulse-level access to ion-trap devices on the cloud to the best of our knowledge, 
this AAIS can be implemented on the IBM's device~\cite{ibm_quantum_2021} and IonQ's device~\cite{chen2024benchmarking}. 
%However, regular could users currently do not have direct low-level access to tune those amplitudes.

\subsection{Related Work and Basic Compilation Process} \label{Sec:CompilationProcess}
Most quantum compilation efforts~\cite{bassman2022arqtic, li2022paulihedral, van2020circuit} regarding quantum simulation is for digital quantum simulation where the compilation for analog quantum simulation is less visited.
There exist several software frameworks~\cite{cross2022openqasm, bloqade, silverio2022pulser} that support direct pulse-level programming, but SimuQ~\cite{peng2024simuq} is the only one that has compilation support.
We now introduce the basics of the compilation for analog quantum simulators. Our introduction will focus on the target quantum system with a time-independent Hamiltonian. Systems with time-dependent Hamiltonians can be discretized in the time axis and then approximated using piece-wise time-independent Hamiltonians~\cite{peng2024simuq}. Thus, accommodating time-independent Hamiltonian can be easily extended to include time-dependent cases. 

When the system Hamiltonian is time-independent, the evolution of a quantum system is given by:
\begin{equation*}
    \lvert \Psi(t) \rangle = e^{-i\frac{H}{\hbar}t} \lvert \Psi(0) \rangle.
\end{equation*}
Then compiling a target quantum system to a quantum simulator is to adjust the variables and evolution time of the simulator such that:
\begin{equation}
    H_{sim} \times T_{sim} = H_{tar} \times T_{tar},
    \label{eqs:match}
\end{equation}
where $T_{sim}$ and $T_{tar}$ are the simulator evolution time and target evolution time, respectively, and we call $T_{sim}$ the \textbf{evolution time variable}.
Notice that these Hamiltonians are usually constructed by Hamiltonian terms, 
\begin{equation}
    H_{sim} = \sum_{i=1}^{L} A^{i}_{sim} H_{i}, \enspace H_{tar} = \sum_{i=1}^{L} A^{i}_{tar} H_{i}, 
    \label{eqs:HamDecomp}
\end{equation}
the Equation (\ref{eqs:match}) can be reduced to:
\begin{equation}
    B^{i}_{sim}   \triangleq  A^{i}_{sim} \times T_{sim} = B^{i}_{tar} \triangleq A^{i}_{tar} \times T_{tar} ,
    \label{eqs:matchamp}
\end{equation}
for each Hamiltonian term $H_i$.

Compiler SimuQ~\cite{peng2024simuq} formulates this problem into solving a mixed equation system. We introduce the process using the following three-qubit Ising chain model as an example with $T_{tar}=1$. The target Hamiltonian is:
\begin{equation*}
    H_{tar} = Z_{1}Z_{2} + Z_{2}Z_{3} + X_{1} + X_{2} + X_3.
\end{equation*}
%In the rest of this paper, we will mostly use the
Suppose the Rydberg AAIS is the simulator's instruction set (the compilation for the Heisenberg AAIS will be similar).
For each Hamiltonian terms in the $H_{tar}$, we construct an equation according to Equation (\ref{eqs:matchamp}):
\begin{align*}
    & \frac{C_6}{4|x_1 - x_2|^6} \times T_{sim} = 1  \times 1 \enspace (Z_{1}Z_{2}), \\
    & \frac{C_6}{4|x_2 - x_3|^6} \times T_{sim} = 1  \times 1 \enspace (Z_{2}Z_{3}), \\
% \end{align*}
% \vspace{-36pt}
% \begin{align*}
    & S_{\Omega_1} \times \frac{\Omega_1}{2} \cos\left(\phi_1\right) \times T_{sim} = 1 \times 1 \enspace (X_{1}),  \\
    & S_{\Omega_2} \times \frac{\Omega_2}{2} \cos\left(\phi_2\right) \times T_{sim} = 1 \times 1 \enspace (X_{2}), \\
    & S_{\Omega_3} \times \frac{\Omega_3}{2} \cos\left(\phi_3\right) \times T_{sim} = 1 \times 1 \enspace (X_{3}),    
\end{align*}
where $S_{\Omega_i}$ is an \textbf{indicator variable} that takes value in $\{0, 1\}$, means the on and off for runtime dynamic variables. However, it can be absorbed into the amplitude variable $\Omega_{i}$. 
In addition, the $\hat{n}_i \hat{n}_j$ will introduce extra $Z_{i}$ terms, the Rabi drive $\Omega_i$ will introduce extra $Y_{i}$ terms, and term $Z_{3}Z_{1}$ is not in the target Hamiltonian, we should also include in the equation system:
\begin{align*}
    & \frac{C_6}{4|x_3 - x_1|^6} \times T_{sim} = 0 \enspace (Z_{3}Z_{1}),  \\
    & \left(- \frac{C_6}{4|x_3 - x_1|^6} - \frac{C_6}{4|x_1 - x_2|^6} +   \frac{\Delta_1}{2}\right) \times T_{sim} = 0 \enspace (Z_{1}),  \\
    & \left(- \frac{C_6}{4|x_1 - x_2|^6} - \frac{C_6}{4|x_2 - x_3|^6} +   \frac{\Delta_2}{2}\right) \times T_{sim} = 0 \enspace (Z_{2}),  \\
    & \left(- \frac{C_6}{4|x_2 - x_3|^6} - \frac{C_6}{4|x_3 - x_1|^6} +   \frac{\Delta_3}{2}\right) \times T_{sim} = 0 \enspace (Z_{3}),  \\
% \end{align*}
% \vspace{-36pt}
% \begin{align*}
    & - \frac{\Omega_1}{2} \sin\left(\phi_1\right) \times T_{sim} = 0  \enspace (Y_{1}),  \\
    & - \frac{\Omega_2}{2} \sin\left(\phi_2\right) \times T_{sim} = 0  \enspace (Y_{2}), \\
    & - \frac{\Omega_3}{2} \sin\left(\phi_3\right) \times T_{sim} = 0  \enspace (Y_{3}).  
\end{align*}
SimuQ constructs the above mixed equation system and utilizes SciPy~\cite{virtanen2020scipy} to solve it.
Finally, by setting up the simulator according to the solutions, the evolution of target system can be simulated.

\iffalse
\subsection{Advantage of Analog Quantum Simulation}\label{Sec:AdvantageofAnalogQuantumSimulation}
Compared with digital quantum simulation~\cite{georgescu2014quantum}, where quantum gates are used to construct the quantum circuit, analog quantum simulation offers distinct benefits particularly in scenarios where the system being studied closely resembles the quantum simulator’s natural dynamics. 

Even for a small quantum system, digital quantum simulation requires millions of quantum operations to accurately decompose the unitary evolution, and even more when employing fault-tolerant schemes. Such an immense number of operations is not expected to be feasible in the near future. In contrast, an analog quantum simulator can naturally evolve under specific pulse controls, enabling a more direct simulation approach. This approach enables higher-fidelity modeling of large and complex quantum systems, offering a compelling alternative to digital quantum simulation for scientific exploration. 
\jz{this is overlapped with introduction}

\fi
\section{Limitation of State-of-The-Art} \label{Sec:ProblemFormulation}
In this section, we summarize two key limitations of the state-of-the-art compiler SimuQ~\cite{peng2024simuq} for analog quantum simulation. 

%This section will analyze the limitation of current 
 %compiler SimuQ and formulate the optimization opportunities to overcome these challenges. 

 %\subsection{Limitations of SimuQ} \label{Sec:LimitationsofSimuQ}
 First, we observe that the \textbf{compilation time} is significantly long for SimuQ due to the need to solve a global mixed equation system introduced in Section~\ref{Sec:CompilationProcess}. This complexity arises in advanced optimization stages, where the equation solver must analyze intricate dependencies between amplitude variables, evolution time variables, and indicator variables. The mixed nature of these equations, combining both discrete and continuous variables, leads to a combinatorial explosion in the solution space, resulting in substantial compilation overhead. Thus, SimuQ cannot efficiently tackle the compilation tasks when we increase the size of the target quantum system. 
 We tested SimuQ on compiling the Ising cycle model of different sizes. 
 Its compilation, as shown in Table~\ref{tab:SimuQCompilationTime},  grows exponentially as the number of qubits increases.
 
 %we provide compilation time for different size Ising cycle model in Table~\ref{tab:SimuQCompilationTime}, which grows exponentially as we add more qubits to the system.

 \begin{table}[h]
  \centering
%   \vspace{-5pt}
  \caption{SimuQ compilation time for Ising cycle}
 % \vspace{-5pt} 
    \begin{tabular}{|c|c|c|c|c|c|}
    \hline
    Qubit\#  & 20 & 40 & 60 & 80 & 100 \\
   \hline
   Compilation Time(s) & 11 & 325 & 2111 & 8695 & 23902 \\
    \hline
    \end{tabular}%
 %   \vspace{-10pt}
  \label{tab:SimuQCompilationTime}%
\end{table}%

%\jz{
Second, the \textbf{compiled machine evolution time} is often suboptimal. Recalling the compilation process discussed in Section~\ref{Sec:CompilationProcess}, solving such an equation system typically involves imposing constraints on variables, such as the maximum amplitude and the upper limit on the simulator’s evolution time. The equation solver then provides a feasible solution but does not necessarily find the one that minimizes the evolution time on the simulator. Moreover, based on our experience with SimuQ, the resulting evolution time is often far from optimal. One might suggest setting all pulses to their maximum rate to achieve the shortest possible evolution time. However, in practice, not all instructions will evolve at their maximum capacity, making this approach impractical.
\section{Equation System Optimization} \label{Sec:EquationSystemOptimization}
%This section presents the equation system optimization technique in 
To tackle the challenges above, we propose 
\myCompilerName, a robust and efficient compiler for analog quantum simulation. 
In this section, we will introduce our optimization on handling the global mixed equation system. 
Our key idea is to design a two-level equation solving process. %in the compilation process.
Instead of directly solving the complex mixed equation system, \myCompilerNameSpace will convert the global mixed equation system into a \textit{global linear equation system} and multiple \textit{localized mixed equation systems}, connected by new synthesized intermediate variables.
\myCompilerNameSpace first determines the synthesized intermediate variables by solving the global linear equation system, then computes the amplitude variables by solving the smaller localized mixed equation systems, utilizing the known solutions of the intermediate variables.
This decomposition significantly enhances the efficiency of the equation-solving process.

%The approach involves decomposing the complete mixed equation system into a global linear equation system and multiple localized mixed equation systems. This decomposition significantly enhances the efficiency of the equation-solving process.

\subsection{Global Linear Equation System Builder} \label{Sec:LinearEquationSystemBuilder}
\begin{figure*}[t]
    \centering
 %   \vspace{3pt}
    \includegraphics[width=1\textwidth]{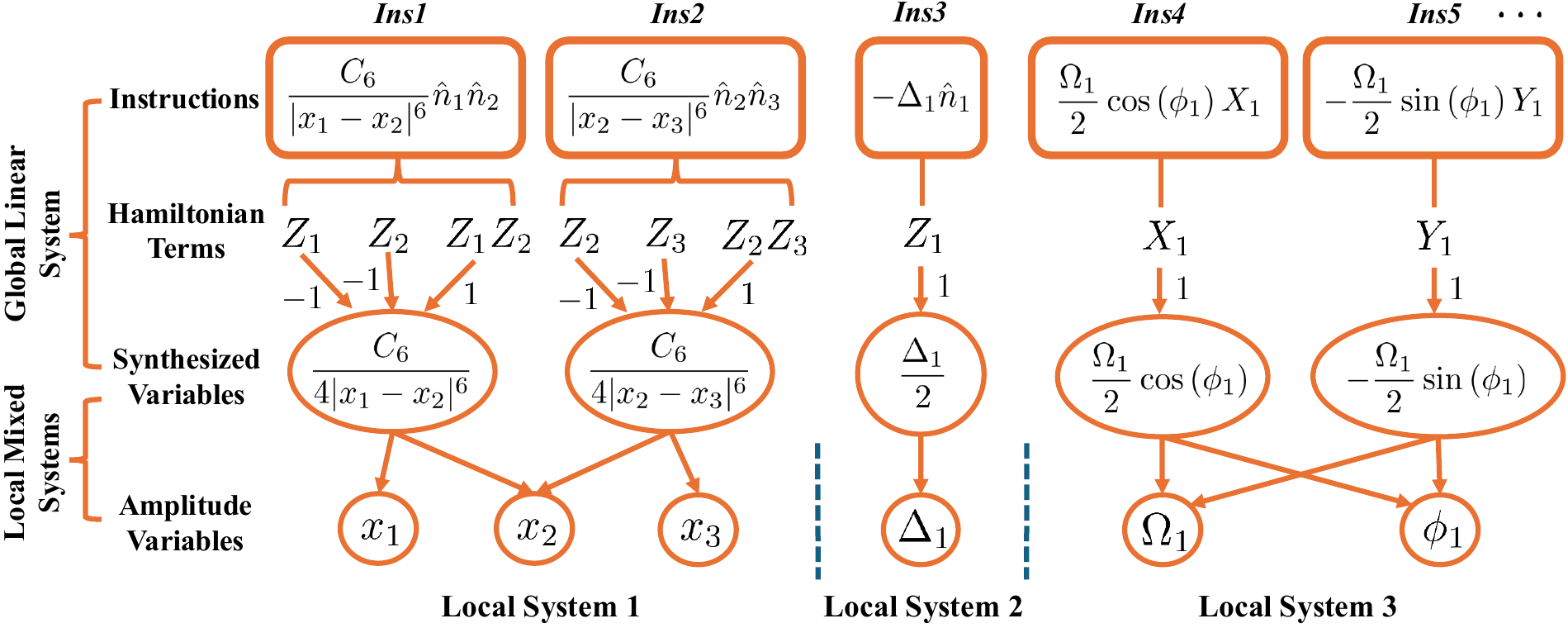}
   % \vspace{-13pt}
    \caption{Process of building the global linear system and local mixed system} %ensen-Shannon Divergence\TODO{check, legend, label}}
%    \vspace{-12pt}
    \label{fig:VarSyn}
    \Description{}
\end{figure*}

We first introduce how to simplify the mixed equation system into a global linear equation system using an example in Fig. \ref{fig:VarSyn}.
In the rest of our introduction to \myCompilerName, we assume that the underlying analog simulator has the Rydberg AAIS as it is more complicated.

In the “Instructions” layer, a subset of the AAIS for a Rydberg device is presented. The first two instructions correspond to Van der Waals interactions between qubits 1 and 2, and qubits 2 and 3, respectively. They directly affects multiple Hamiltonian terms. For instance, the instruction:
\begin{equation*}
    \frac{C_6}{|x_1 - x_2|^6}\hat{n}_1 \hat{n}_2 =  \frac{C_6}{|x_1 - x_2|^6} \times \frac{I -Z_1-Z_2+Z_1Z_2}{4}
\end{equation*}
affects the strengths of the Hamiltonian terms $Z_{1}$, $Z_{2}$, and $Z_{1}Z_{2}$ (as shown in ``Hamiltonian Terms'' layer in Fig. \ref{fig:VarSyn}). The corresponding amplitudes are $-C_6/(4|x_1 - x_2|^6)$, $-C_6/(4|x_1 - x_2|^6)$, and $C_6/(4|x_1 - x_2|^6)$ respectively. We notice that the term $C_6 / (4|x_1 - x_2|^6)$ appears frequently in the mixed equation system (in Section~\ref{Sec:CompilationProcess}), it can be substituted with a new variable, referred to as a synthesized variable (as shown in ``Synthesized Variables'' layer in Fig. \ref{fig:VarSyn}), with direct arrows and associated coefficients indicating the dependency of the Hamiltonian terms on the synthesized variable.
The analysis for other instructions
%\begin{equation*}
%   -\Delta_i\hat{n}_i, \hspace{0.2em} \frac{\Omega_i}{2} \cos\left(\phi_i\right) X_i, \hspace{0.2em} and \enspace -\frac{\Omega_i}{2} \sin\left(\phi_i\right) Y_i
%\end{equation*}
is a similar approach, and the results are on the right of Fig. \ref{fig:VarSyn}.

At this stage, one could substitute the synthesized variables into the mixed equation system. However, we observe that in the mixed equation system, these variables are always multiplied by the evolution time variable $T_{sim}$. 
% \jz{The product of the amplitude and the evolution time forms a scaler without unit. }
Consequently, %rather than introducing the synthesized variables directly as new variables in the linear system, 
we define the product of the synthesized variables and the evolution time as new synthesized variables. This approach ensures that once an evolution time is determined, the corresponding values of the synthesized variables can be easily calculated.

We now proceed to construct the global linear equation system, using the three-qubit Ising chain model in Section~\ref{Sec:CompilationProcess} as an example. The variable substitution is performed as follows. For Van der Waals interactions:
\begin{equation}
    \left\{
\begin{aligned}
    & \frac{C_6}{4|x_1 - x_2|^6} \times T_{sim} \xrightarrow[]{} \alpha_1, \\
    & \frac{C_6}{4|x_2 - x_3|^6} \times T_{sim} \xrightarrow[]{} \alpha_2, \\
    & \frac{C_6}{4|x_3 - x_1|^6} \times T_{sim} \xrightarrow[]{} \alpha_3.\\
\end{aligned}
\right.
\label{eq:possub}
\end{equation}
For detuning instructions:
\begin{equation*}
    \left\{
\begin{aligned}
    & \frac{\Delta_1}{2} \times T_{sim} \xrightarrow[]{} \alpha_4, \\
    & \frac{\Delta_2}{2} \times T_{sim} \xrightarrow[]{} \alpha_5, \\
    & \frac{\Delta_3}{2} \times T_{sim} \xrightarrow[]{} \alpha_6.\\
\end{aligned}
\right.
\end{equation*}
For Rabi drive instructions:
\begin{equation*}
    \left\{
\begin{aligned}
    & \frac{\Omega_1}{2} \cos\left(\phi_1\right) \times T_{sim} \xrightarrow[]{} \alpha_7, \enspace -\frac{\Omega_1}{2} \sin\left(\phi_1\right) \times T_{sim} \xrightarrow[]{} \alpha_8, \\
    & \frac{\Omega_2}{2} \cos\left(\phi_2\right) \times T_{sim} \xrightarrow[]{} \alpha_9, \enspace -\frac{\Omega_2}{2} \sin\left(\phi_2\right) \times T_{sim} \xrightarrow[]{} \alpha_{10}, \\
    & \frac{\Omega_3}{2} \cos\left(\phi_3\right) \times T_{sim} \xrightarrow[]{} \alpha_{11}, \enspace -\frac{\Omega_3}{2} \sin\left(\phi_3\right) \times T_{sim} \xrightarrow[]{} \alpha_{12}.\\
\end{aligned}
\right.
\end{equation*}
Using these substitutions, the global linear equation system upon the synthesized variables is constructed as:
\begin{equation}
    \left\{
\begin{aligned}
    & \alpha_1 = 1, \enspace \alpha_2 = 1, \enspace \alpha_3 = 0, \\
    & -\alpha_{1} - \alpha_{3} + \alpha_{4} = 0,\\
    & -\alpha_{1} - \alpha_{2} + \alpha_{5} = 0,\\
    & -\alpha_{2} - \alpha_{3} + \alpha_{6} = 0,\\
    & \alpha_{7} = 1, \enspace \alpha_{9} = 1, \enspace \alpha_{11} = 1, \\
    & \alpha_{8} = 0, \enspace \alpha_{10} = 0, \enspace \alpha_{12} = 0. \\
\end{aligned}
\right.
\label{eq:linsys}
\end{equation}
This linear system can be efficiently solved, yielding the solution:
\begin{equation*}
    \left\{
\begin{aligned}
    & \alpha_1 = 1, \enspace \alpha_2 = 1, \enspace \alpha_3 = 0, \\
    &  \alpha_4 = 1, \enspace \alpha_5 = 2, \enspace \alpha_6 = 1, \\
    & \alpha_{7} = 1, \enspace \alpha_{9} = 1, \enspace \alpha_{11} = 1, \\
    & \alpha_{8} = 0, \enspace \alpha_{10} = 0, \enspace \alpha_{12} = 0. \\
\end{aligned}
\right.
\end{equation*}

\subsection{Localized Mixed Equation System Builder}\label{Sec:LocalizedMixedEquationSystemBuilder}

In the previous section, we formulated a global linear equation system to align the simulator's evolution with the target system. However, its solution, expressed in synthesized variables, is only an intermediate result. To fully resolve the problem, we must determine the amplitude variables and the evolution time variable.

To achieve this, we need to first identify the dependencies between different synthesized variables, enabling us to solve them independently. In the lower half of Fig. \ref{fig:VarSyn}, direct arrows illustrate the connections between each synthesized variable and its corresponding amplitude variables. For instance, the synthesized variable $C_6/4|x_1 - x_2|^6$ depends on the amplitude variables $x_1$ and $x_2$, while $C_6/4|x_2 - x_3|^6$ depends on $x_2$ and $x_3$. These two synthesized variables are ``connected'' through the shared amplitude variable $x_2$. In contrast, they are not connected to $\Delta_1 /2$, as detuning instructions do not depend on atom positions.

We can now reformulate the problem of identifying dependencies as a graph problem. In this representation, the synthesized variables and the amplitude variables serve as nodes, while the arrows between them correspond to edges. The problem then reduces to identifying the connected components within this graph.

Once the variable connections are identified, we can decompose the entire system into smaller, localized subsystems and solve each localized equation system independently. As illustrated in Fig. \ref{fig:VarSyn}, the presenting system is partitioned into three distinct local components labeled by ``Local System 1/2/3''.

As a result, we can solve position variables $x_{1}$, $x_{2}$, and $x_{3}$ together. Since they contribute to the synthesized variables $\alpha_{1}$, $\alpha_{2}$, and $\alpha_{3}$ in the linear system, the localized mixed equation is constructed as:
\begin{equation*}
    \left\{
\begin{aligned}
    & \frac{C_6}{4|x_1 - x_2|^6} \times T_{sim} = 1, \\
    & \frac{C_6}{4|x_2 - x_3|^6} \times T_{sim} = 1, \\
    & \frac{C_6}{4|x_3 - x_1|^6} \times T_{sim} = 0.\\
\end{aligned}
\right.
\end{equation*}
The Rabi drive variables $\Omega_{1}$ and $\phi_{1}$ are connected, and they contribute to the synthesized variables $\alpha_{7}$, $\alpha_{8}$ in the linear system. The localized mixed equation is constructed as:
\begin{equation*}
    \left\{
\begin{aligned}
    & \frac{\Omega_1}{2} \cos\left(\phi_1\right) \times T_{sim} =1, \\
    & \frac{\Omega_1}{2} \sin\left(\phi_1\right) \times T_{sim} =0.\\
\end{aligned}
\right.
\end{equation*}
The detuning parameter $\Delta_{1}$ itself forms a isolated variable, and the corresponding synthesized variable in linear system is $\alpha_{4}$. The localized mixed equation is:
\begin{equation*}
    \frac{\Delta_1}{2} \times T_{sim} = 1
\end{equation*}

By identifying these connected components, we can systematically construct localized mixed equation systems for each group.
Compared with the global mixed system employed by SimuQ~\cite{peng2024simuq}, the localized equation system in \myCompilerNameSpace is a more efficient and targeted approach to solving for the amplitude variables and evolution time variable.

\section{ Evolution Time Optimization} \label{Sec:EvolutionTimeOptimization}

In the previous section, we constructed a global linear system and utilized its solution to derive multiple localized mixed equation systems, whose solutions constitute the final result. In this section, we detail the methodology for determining the evolution time and amplitude variables on a quantum device using these localized mixed equation systems. The proposed strategy involves solving the localized mixed equation systems for runtime dynamic variables and runtime fixed variables separately.

\subsection{Finding Optimized Evolution Time Using Runtime Dynamic Variables} \label{Sec:FindingOptimizedEvolutionTime UsingLocalVariables}
The runtime dynamic variables, as introduced in Section~\ref{Sec:AnalogQuantumSimulation}, are parameters whose values can change dynamically during program execution. This section outlines the method for leveraging the constraints imposed on these runtime dynamic variables to determine a near-optimal machine evolution time for the analog quantum simulator.

From this point, we introduce an assumption regarding the instructions in AAIS controlled by runtime dynamic variables: each instruction contains a variable that directly governs the amplitude of Hamiltonian terms, referred to as the \textbf{time-critical variable}. This variable directly impacts the amplitude scaling of the Hamiltonian terms. For instance, in the Rydberg AAIS, the time-critical variable is $\Delta$ for the detuning instruction and $\Omega$ for the Rabi drive. In the Heisenberg AAIS, all variables $a^{P_i}$ and $a^{P_iP_j}$ are classified as time-critical variables. 
%\todo{a new term again...time-critical variable....}

The rationale behind optimizing evolution time lies in the varying upper bounds of different time-critical variables. At maximum capability, instructions with runtime dynamic variables may reach the solution derived from the global linear system at different rates. The critical aspect of this optimization is identifying the slowest-evolving instruction, which serves as the bottleneck in the evolution process. The evolution time of this bottleneck instruction, when operating at its maximum capacity, is then set as the overall simulator evolution time. This approach ensures that the bottleneck is fully utilized while keeping all other instructions within a safe operational range.

We continue our analysis using the three-qubit Ising chain model introduced in Section~\ref{Sec:EquationSystemOptimization}, examining the case-by-case behavior. The constraints on the following time-critical variable in the rest of this section are set based on the specification of QuEra's Aquila device~\cite{wurtz2023aquila}.

\textbf{Case 1:} for the first detuning instruction, the localized mixed equation system is given as:
\begin{equation*}
    \frac{\Delta_1}{2} \times T_{sim} = 1,
\end{equation*}
where $\Delta_1$ has a maximum amplitude of $20 \hspace{2pt} MHz$. The shortest evolution time for this instruction can be calculated as:
\begin{equation*}
    T_{sim} = \frac{2}{20*10^6} \hspace{2pt} s=0.1 \hspace{2pt} \mu s.
\end{equation*}
Similarly, the shortest evolution times for the second and the third detuning instructions are calculated as $0.2 \hspace{2pt} \mu s$ and $0.1 \hspace{2pt} \mu s$, respectively.

\textbf{Case 2:} for the first Rabi drive instruction, the localized mixed equation system is given as:
\begin{equation*}
    \left\{
\begin{aligned}
    & \frac{\Omega_1}{2} \cos\left(\phi_1\right) \times T_{sim} =1, \\
    & \frac{\Omega_1}{2} \sin\left(\phi_1\right) \times T_{sim} =0.\\
\end{aligned}
\right.
\end{equation*}
In this case, we first absorb the evolution time into the time-critical variable $\Omega_1$ as a single term and solve the system as:
\begin{equation*}
    \Omega_1 \times T_{sim} = 2, \enspace \phi_1 = 0.
\end{equation*}
Using the maximum amplitude of $\Omega_1$, which is $2.5 \hspace{2pt} MHz$ , the shortest evolution time for this instruction is calculated as:
\begin{equation}
    T_{sim} = \frac{2}{2.5*10^6}\hspace{2pt} s=0.8 \hspace{2pt} \mu s.
    \label{eq:evosol}
\end{equation}
The shortest evolution times for the second and third Rabi drive instructions are identical to this value.

\textbf{Case 3:} an additional scenario, which is not present in the previous examples, arises when there is no time-critical variable in the localized mixed equation system. For instance, consider the following case:
\begin{equation*}
    \cos\left(\phi\right) \times T_{sim} =1.
\end{equation*}
In such a situation, the absence of an time-critical variable can be addressed by introducing an additional constraint:
\begin{equation*}
    minimize \enspace T_{sim}.
\end{equation*}
This approach allows us to determine the shortest evolution time by solving the system under the given constraints. The result of given problem is $\phi = 0$ and $T=1$

After determining the evolution times for each instruction, we select the longest one, $T_{sim} =0.8 \hspace{2pt} \mu s$, as the simulator evolution time. Using this updated evolution time, we reconfigure the previous localized equation systems and obtain the following solutions:
\begin{equation*}
    \Delta_1 = \Delta_3 = 2.5 \hspace{2pt} MHz, \enspace \Delta_2 = 5.0 \hspace{2pt} MHz, 
\end{equation*}
and:
\begin{equation*}
    \Omega_1 = \Omega_2 = \Omega_3 = 2.5 \hspace{2pt} MHz, \enspace \phi_1 = \phi_2 = \phi_2 = 0.
\end{equation*}
All these values satisfy the hardware constraints, ensuring that the bottleneck—the Rabi drive—is operating at its maximum capability without exceeding any limitations.

\subsection{Runtime Fixed Variables Solver} \label{Sec:GlobalVariablesSolver}

With the evolution time determined, we can now proceed to solve the localized mixed equation systems for the runtime fixed variables. These variables are fixed and cannot be altered once the program execution has commenced.

By directly substituting $T_{sim}$ into the equations, we derive the mixed equation system for the runtime fixed variables:
\begin{equation}
    \left\{
\begin{aligned}
    & \frac{C_6}{4|x_1 - x_2|^6} = 1.25, \\
    & \frac{C_6}{4|x_2 - x_3|^6} = 1.25, \\
    & \frac{C_6}{4|x_3 - x_1|^6} = 0.\\
\end{aligned}
\label{eq:globalsolver}
\right.
\end{equation}
The solution to this system is:
\begin{equation}
    x_{1} =  0 \hspace{2pt} \mu m, \enspace x_{2} =  7.46 \hspace{2pt} \mu m, \enspace x_{3} =  14.92 \hspace{2pt} \mu m.
    \label{eq:postionsolution}
\end{equation}
Note that in practice, atom position variables are two-dimen-sional vectors. Here, we consider a one-dimensional scalar case for simplicity.

It is important to note that runtime fixed variables like this position of atoms may be subject to specific constraints. If the solution derived using the optimal evolution time from Section~\ref{Sec:FindingOptimizedEvolutionTime UsingLocalVariables} does not satisfy these constraints, an iterative approach can be employed. This involves incrementally increasing the evolution time by $\Delta t$ and recalculating the solution until the constraints are satisfied. This method balances the trade-off between achieving the shortest possible simulator evolution time and meeting the requirements imposed by the runtime fixed variable constraints. Additionally, this $T_{sim}$ can be utilized to generate a more accurate initial guess for the runtime fixed variables.

\subsection{Solver for Time Dependent Hamiltonian} \label{Sec:SolverforTimeDependentHamiltonian}
In this section, we address the scenario where the Hamiltonian is time-dependent. In such cases, the compilation process differs from that of a time-independent Hamiltonian. To simplify the analysis, we consider a piecewise constant Hamiltonian, as any time-dependent Hamiltonian can be effectively approximated using this approach~\cite{peng2024simuq}.

For each time segment where the Hamiltonian is constant, following the process outlined in the previous two sections, we may derive an equation system similar to Equation (\ref{eq:globalsolver}):
\begin{equation*}
    \left\{
\begin{aligned}
    & \frac{C_6}{4|x_1 - x_2|^6} = \beta_{1,t_{i}}, \\
    & \frac{C_6}{4|x_2 - x_3|^6} = \beta_{2,t_{i}}, \\
    & \frac{C_6}{4|x_3 - x_1|^6} = \beta_{3,t_{i}},\\
\end{aligned}
\right.
\end{equation*}
Here, $\beta_{j,t_i}$ represents the amplitude that the $j$th component in the equation system must achieve during the time segment $[t_{i},t_{i+1}]$. Importantly, for all $t_{i}$, the amplitude $\beta_{j,t_i}$ can be lower but cannot exceed its specified value, as it corresponds to the shortest simulator evolution time for that segment.

To address this, we identify the smallest set of $\{\beta_{j,t_i}\}_{j=1}^{m}$ across the time segment $[t_{i},t_{i+1}]$ and use its corresponding solution for the runtime fixed variables to reconfigure the simulator evolution time for other time segments. This approach inherently increases the simulator evolution time for those segments, thereby lowering the amplitude of each time-critical variable to match the target evolution as the simulator evolution time is extended. By ensuring that the amplitudes are reduced when the evolution time is increased, this method guarantees that the solutions always comply with hardware constraints.

\section{Accuracy Control} \label{Sec:AccuracyControl}

In the previous two sections, we discussed the decomposition of the mixed equation system into a global linear system and several localized mixed equation systems, allowing us to efficiently obtain solutions. However, there are cases where we cannot get the exact solution and can only approximate the target evolution.
In this section, we analyze how errors propagate throughout this procedure and introduce an optimization method designed to enhance the approximation accuracy. 

\subsection{Compilation Approximation Error Analysis} \label{Sec:ErrorAnalysis}
We first define a metric to quantify the total error in the compilation process. As described in Section~\ref{Sec:CompilationProcess}, the goal is to match $H_{sim} \times T_{sim}$ with $H_{tar} \times T_{tar}$, by combining Equation (\ref{eqs:HamDecomp}) and (\ref{eqs:matchamp}), they can be expressed as:
\begin{equation*}
    H_{sim} \times T_{sim} = \sum_{i=1}^{L} B^{i}_{sim} H_{i}, \enspace H_{tar} \times T_{tar} = \sum_{i=1}^{L} B^{i}_{tar} H_{i}.
\end{equation*}
By defining $B_{sim}$ as the vector containing the coefficients $B^{i}_{sim}$ and $B_{tar}$ as the vector containing $B^{i}_{tar}$, we can use the following metric to capture the difference between the two evolutions:
\begin{equation}
    E \triangleq||B_{sim}-B_{tar}||_{1}
    \label{eq:errormetric}
\end{equation}
where $||\cdot||_1$ represents the $L_1$-norm. This metric provides a straightforward measure of the discrepancy between the simulated and target evolutions. We now provide the error bound of compilation process.

\begin{theorem} \label{Th:errorbound}
Suppose the $L_1$ error in solving the global linear equation system is bounded by $\epsilon_1$, and the $L_1$ error in solving each localized mixed equation system is bounded by $\epsilon_2^i$, then the total error in Equation (\ref{eq:errormetric}) is bounded by:
\begin{equation}
    \vert \vert M \vert \vert_{1} \cdot \sum_{i=1}^{K}\epsilon_2^i + \epsilon_1,
    \label{eq:errorbound}
\end{equation}
where $M$ is the matrix that governing the global linear system and $K$ is the number of localized mixed equation systems.

\begin{proof}
    The proof is in Appendix~\ref{sec:appendix}.
\end{proof}

\end{theorem}

This error bound provides a structured framework for understanding the propagation of errors across both the global linear system and the localized mixed equation systems. This implies that if we can ensure the errors in the global linear system and the localized mixed systems are sufficiently small, we can efficiently approximate the target quantum system’s evolution with high accuracy.

\subsection{Enhanced Optimization for Accuracy Improvement} \label{Sec:EnhancedOptimizationforAccuracyImprovement}

The solution obtained by first solving a global linear system and subsequently addressing several localized mixed equation systems may not be globally optimal. To illustrate this, we continue analyzing the three-qubit Ising chain model discussed in the previous section.

By substituting the solution for the position variables from Equation (\ref{eq:postionsolution}) and the evolution time solution from Equation (\ref{eq:evosol}) into Equation (\ref{eq:possub}), we can compute the values of the synthesized variables:
\begin{equation*}
    \alpha_{1} = 1.001, \enspace \alpha_{2} = 1.001, \enspace \alpha_{3} = 0.020.
\end{equation*}
Next, inserting these values into a portion of the linear equation system shown in Equation (\ref{eq:linsys}):
\begin{equation*}
    \left\{
\begin{aligned}
    & -\alpha_{1} - \alpha_{3} + \alpha_{4} = 0,\\
    & -\alpha_{1} - \alpha_{2} + \alpha_{5} = 0,\\
    & -\alpha_{2} - \alpha_{3} + \alpha_{6} = 0,\\
\end{aligned}
\right.
\end{equation*}
we update the values for $\alpha_{4}$, $\alpha_{5}$, and $\alpha_{6}$ as follows: 
\begin{equation*}
    \alpha_{4} = 1.021, \enspace \alpha_{5} = 2.002, \enspace \alpha_{6} = 1.021.
\end{equation*}
These updated values correspond to the following detuning amplitudes:
\begin{equation*}
    \Delta_1 = \Delta_3 = 2.55 \hspace{2pt} MHz, \enspace \Delta_2 = 5.01 \hspace{2pt} MHz.
\end{equation*}
This result remains within the allowable range for detuning amplitudes. Furthermore, this update enhances the accuracy of the solution, demonstrating the benefit of iterative refinement in this approach.

We now introduce a systematic approach to optimize the solution through only one iterative refinement.
Suppose we decompose the global linear matrix $M$ into parts that govern synthesized runtime fixed variables $\alpha_{r}$ and runtime dynamic variables $\alpha_{c}$ seperately:
\begin{equation*}
    M = \begin{pmatrix}
M_r & M_c 
\end{pmatrix}.
\end{equation*}
The global linear equation system then becomes:
\begin{equation*}
\begin{pmatrix}
M_r & M_c 
\end{pmatrix}
\begin{pmatrix}
\alpha_{r} \\
\alpha_{c}
\end{pmatrix}
=
\begin{pmatrix}
B_{tar}
\end{pmatrix}.
\end{equation*}
After solving the system in the first round, we obtain an approximate solution:
\begin{equation*}
\begin{pmatrix}
M_r & M_c 
\end{pmatrix}
\begin{pmatrix}
\widetilde{\alpha}_g \\
\widetilde{\alpha}_l
\end{pmatrix}
=
\begin{pmatrix}
\widetilde{B}_{tar}
\end{pmatrix}.
\end{equation*}
By comparing the difference between the exact and approximate equations, we derive the residual equation: 
\begin{equation*}
\begin{pmatrix}
M_r & M_c 
\end{pmatrix}
\begin{pmatrix}
\delta \alpha_{r} \\
\delta \alpha_{c}
\end{pmatrix}
=
\begin{pmatrix}
\delta B_{tar}
\end{pmatrix},
\end{equation*}
where $\delta$ represents the residuals. This leads to the expression:
\begin{equation*}
    M_r \cdot \delta \alpha_{r} + M_c \cdot \delta \alpha_{c} = \delta B_{tar}.
\end{equation*}
Ideally, the residual term $\delta B_{tar}$ should be as small as possible, as it represents the error in the equation system. Notably, since the runtime dynamic variables can be adjusted during program execution and are more flexible than runtime fixed variables, we can minimize:
\begin{equation*}
    ||M_r \cdot \delta \alpha_{r} + M_c \cdot \delta \alpha_{c}||_{1}
\end{equation*}
by treating $\delta \alpha_{c}$ as the optimization variable and modifying the corresponding instructions accordingly. This adjustment refines the solution iteratively, thereby reducing the error in the overall solution.
\section{Evaluation} \label{Sec:Evaluation}
In this section, we evaluate \myCompilerNameSpace against state-of-the-art baseline (SimuQ~\cite{peng2024simuq}) in terms of compilation time, execution time, and program accuracy. Additionally, we conduct experiments on a real quantum device to assess how \myCompilerNameSpace enhances the robustness of analog quantum programs.

\subsection{Experiments Setup} \label{Sec:ExperimentsSetup}

\textbf{Baseline:} Our baseline compiler is SimuQ~\cite{peng2024simuq}, which is the only publicly available compiler for analog quantum simulators to the best of our knowledge.

\textbf{Aanlog Simulator Backend} We employ the Rydberg AAIS~\cite{wurtz2023aquila} and Heisenberg AAIS~\cite{chen2024benchmarking, ibm_quantum_2021} as introduced in Section~\ref{Sec:AnalogQuantumSimulation}. The physical pulse constraints of Rydberg AAIS and Heisenberg AAIS are set according to ~\cite{wurtz2023aquila} and ~\cite{kanazawa2023qiskit} respectively.
%\textbf{Experiment Configurations}
%To demonstrate the optimization impact of \myCompilerName, we designed a series of experiments as follows:

%(\textbf{a}) In Section~\ref{Sec:OverallResult}, we conducted an overall experiment involving two simulators—Rydberg AAIS and Heisenberg AAIS—with four target system models assigned to each simulator. This experiment is denoted as “Overall” in Table~\ref{tab:benchmarks}. Each experiment was executed under two configurations: (1) “Baseline,” which employs the sole available analog compiler, SimuQ, and (2) “Opt,” which applies the \myCompilerNameSpace optimization.
%(\textbf{b}) Section~\ref{Sec:MappingandTimeDependentHamiltonian} examines cases that involve site mapping and the handling of time-dependent scenarios on the Rydberg AAIS, using the same configuration as in the overall experiment. These are labeled “Mapping” and “TimeDep” in Table~\ref{tab:benchmarks}, respectively.
%(\textbf{c}) Finally, in Section~\ref{Sec:RealDeviceExecution}, we present real experiments to demonstrate noise mitigation and pulse shrinking, this experiment is referred to as “RealExp” in Table~\ref{tab:benchmarks}.

\textbf{Benchmark} To simulate the target system, we employ a benchmark suite that includes models from various physics domains~\cite{lee1952statistical, takahashi1971one, aramthottil2022scar} as well as optimization~\cite{ebadi2022quantum}. Table~\ref{tab:benchmarks} provides an overview of the benchmarks and their corresponding Hamiltonians (we abbreviate ``Heisenberg chain'' as ``Heis chain'').
%For the simulator, we employ the Rydberg AAIS and Heisenberg AAIS. Table~\ref{tab:benchmarks} presents the benchmark selected for different experiments. 
Specifically, the ``Ising cycle $+$'' model is from ~\cite{dag2024emergent}, the ``PXP'' model is from ~\cite{turner2018quantum}, and all other models can be found in ~\cite{peng2024simuq}. 
Note the ``MIS Chain'' is time-dependent and other benchmarks are time-independent.
All parameters in these target physical model are set to $1 \hspace{2pt} MHz$, and the their evolution time $T_{tar}$ is also fixed at $1 \hspace{2pt} \mu s$, except the real system experiment.
%For the configurations "Overall," "Mapping," and "TimeDep," all parameters of the corresponding physical model are set to 1, and the evolution time is also fixed at 1. The pulse constraints of Rydberg AAIS and Heisenberg AAIS are set according to ~\cite{wurtz2023aquila} and ~\cite{kanazawa2023qiskit} respectively. For the benchmark setting of "RealExp", we will present its details in Section~\ref{Sec:RealDeviceExecution}.

% \begin{table}[t]
%   \centering
%   \caption{Benchmark Information \todo{change this table, just the benchmark names and their Hamiltonian expression, you may also consider put the long experssions in the appendix}}
% %  \vspace{-10pt}
%     \begin{tabular}{|c|c|c|}\hline

% %       Configuration & Rydberg AAIS & Heisenberg AAIS\\
%  %   \hline
%      \multirow{4}{*}{\textbf{Overall}} & Ising chain & Ising chain  \\ 
%      \cline{2-3}
%      & Ising cycle & Ising cycle \\ \cline{2-3}
%      & Kitaev & Kitaev \\ \cline{2-3} 
%      & Ising cycle + & Heis chain \\ \hline
%      {\textbf{Mapping}} & Ising chain & \diagbox{}{}  \\ \hline
%      {\textbf{TimeDep}} & MIS chain & \diagbox{}{}  \\ \hline
%      \multirow{2}{*}{\textbf{RealExp}} & Ising cycle & \diagbox{}{} \\ 
%      \cline{2-3}
%      & PXP & \diagbox{}{} \\ \hline

%     \end{tabular}%
% %    \vspace{-8pt}
%   \label{tab:benchmarks}%
%   \end{table}

  \begin{table}[t]
  \centering
  \renewcommand{\arraystretch}{1.3}
  \caption{Benchmark information. $N$ is the number of qubits.}
%  \vspace{-10pt}
{\fontsize{8.7}{10}\selectfont
    \begin{tabular}{|c|c|}\hline

%       Configuration & Rydberg AAIS & Heisenberg AAIS\\
 %   \hline
     \textbf{Model} & \textbf{Hamiltonian}  \\ \hline
     Ising chain & $ J \cdot \sum_{i=1}^{N-1} Z_{i}Z_{i+1} + h \cdot  \sum_{i=1}^{N} X_{i}$ \\ \hline
     Ising cycle & $ J \cdot \sum_{i=1}^{N} Z_{i}Z_{i+1} + h \cdot  \sum_{i=1}^{N} X_{i}$ \\ \hline
     Kitaev & $ \frac{\mu}{2} \cdot \sum_{i=1}^{N-1} Z_{i}Z_{i+1} -   \sum_{i=1}^{N} (tX_{i}+hZ_{i})$ \\ \hline
      Ising cycle + & $ J \cdot \sum_{i=1}^{N} Z_{i}Z_{i+1} + \frac{J}{2^6} \cdot \sum_{i=1}^{N} Z_{i}Z_{i+2} + h \cdot  \sum_{i=1}^{N} X_{i}$ \\ \hline
      Heis chain & $ J \cdot \sum_{i=1}^{N-1} (X_{i}X_{i+1}+ Y_{i}Y_{i+1} + Z_{i}Z_{i+1}) + h \cdot  \sum_{i=1}^{N} X_{i}$  \\ \hline
    MIS chain & $
    \sum_{i=1}^{N} \left((1-2t)U \hat{n}_i + \frac{\omega}{2} X_i \right) + \sum_{i=1}^{N-1} \alpha \hat{n}_i \hat{n}_{i+1}
    $ \\ \hline
    PXP & $J \cdot \sum_{i=1}^{N-1} n_{i}n_{i+1} + h \cdot \sum_{i=1}^{N} X_{i} = h \cdot \sum_{i=1}^{N} P_{i-1} X_{i} P_{i+1}$ \\ \hline
    \end{tabular}%

    }
%    \vspace{-8pt}
  \label{tab:benchmarks}%
  \end{table}

\textbf{Metrics}
%In the "Overall", "Mapping", and "TimeDep" experiments, three distinct metrics are used to evaluate our compiler's performance. 
We mainly use the following three metrics:
(\textbf{a}) \textbf{Compilation Time:} The CPU time required to compute the pulse solution for a given target system and simulator. (\textbf{b}) \textbf{Execution Time:} The duration required to execute the compiled pulse on a real device. A shorter execution time is preferable, as most current devices are highly sensitive to noise.
(\textbf{c}) \textbf{Program Relative Error:} We define the following ratio as the ``Relative Error'' of the compiled Hamiltonian similar to Equation (\ref{eq:errormetric}) in Section~\ref{Sec:ErrorAnalysis}:
% \begin{equation*}
%    E = \frac{\sum_{i}^L | A^{i}_{sim} \times T_{sim} - A^{i}_{tar} \times T_{tar} |}{\sum_{i}^L |  A^{i}_{tar} \times T_{tar} |} \times 100\%.
% \end{equation*}
\begin{equation*}
   E = \frac{||B_{sim}-B_{tar}||_{1}}{||B_{tar}||_{1}} \times 100\%.
\end{equation*}
It quantifies the discrepancy between the target evolution and the simulator's evolution.
% , where $A^{i}_{sim}$ represents the strength of the simulator's Hamiltonian term, which depends on the solution of the amplitude variables.
%In the "RealExp" experiments, the performance metric is based on practical observables of the quantum system. The time-evolved magnetization and the  domain-wall density are defined as~\cite{dag2024emergent}:
%\begin{equation*}
%   \mathcal{M} = \frac{1}{N} \sum_{i=1}^{N} \langle Z_{i}\rangle, \enspace \mathcal{G} = \frac{1}{N} \sum_{i=1}^{N} \langle Z_{i}Z_{i+1}\rangle.
%\end{equation*} In this context, values obtained from a real device that more closely match the theoretical predictions are considered preferable. 

  \textbf{Implementation} We implemented \myCompilerNameSpace using Python. We used the numpy package~\cite{harris2020array} 
  %is utilized to solve the linear equation system introduced in Section~\ref{Sec:EquationSystemOptimization}, while 
  and the scipy package~\cite{virtanen2020scipy} to solve the equation systems.
  %is employed to address the localized mixed equation system described in Sections~\ref{Sec:EvolutionTimeOptimization} and~\ref{Sec:AccuracyControl}. 
  All programs are executed on a server with CPU Frequency 2.4 GHz and 768 GB memory. For real device evaluation, the quantum program is executed on QuEra’s 256-qubit analog quantum computer Aquila via Amazon Braket.

\subsection{Overall Comparision} \label{Sec:OverallResult}
%In this section, we evaluate our compiler using various simulators and physical models. The "Overall" section in Table~\ref{tab:benchmarks} presents the chosen benchmarks, with physical system sizes ranging from 3 to 100. 

\begin{figure*}[t]
    \centering
 %   \vspace{3pt}
    \includegraphics[width=1\textwidth]{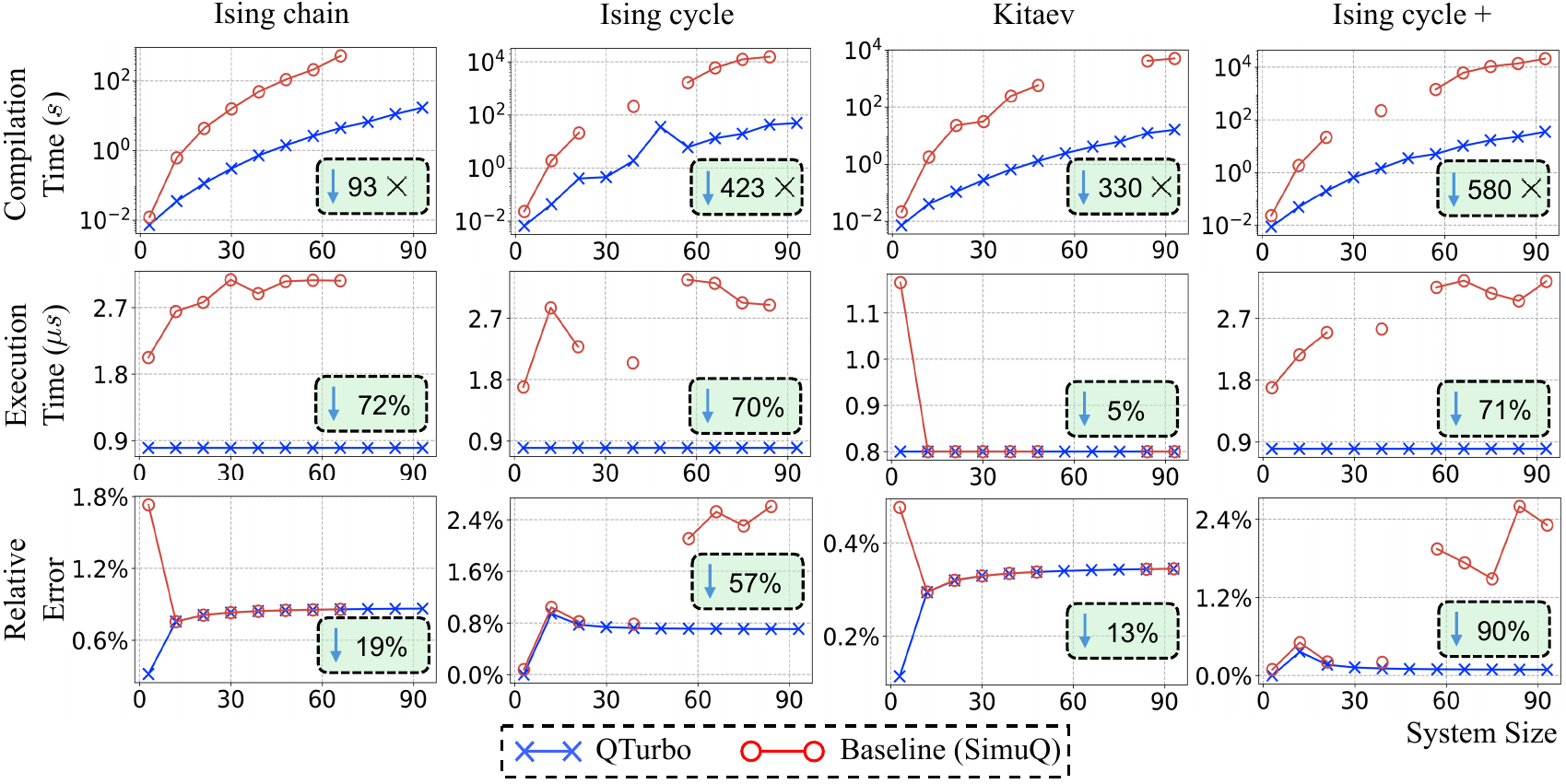}
   % \vspace{-13pt}
    \caption{ Compilation result comparision for Rydberg AAIS} 
  %  \vspace{-12pt}
    \label{fig:CRR}
    \Description{}
\end{figure*}

Figure~\ref{fig:CRR} presents the compilation results for targeting the Rydberg device for four benchmarks (in four columns) of sizes ranging from 3 qubits to 93 qubits. On average, our approach achieves a $350 \times$ speedup in compilation, while reducing execution time and compilation error by $54\%$ and $45\%$, respectively. Occasionally, the SimuQ compiler fails to produce a solution, resulting in missing data points. The detailed breakdown of these findings can be found in the box of each subgraph. 
%The first row, %labeled 'Compilation Time,' 
% The average speedup for individual benchmarks is shown in the boxes in the first row. 
% The average execution time reduction rate and relative error reduction rate in the yellow and green boxes, respectively.
%The second and third rows, 'Execution Time' and 'Relative Error,' provides their reduction rates within the yellow and green boxes respectively. 
Notice that for the ``Kitaev'' model, the SimuQ compiler does find the optimal solution sharing the same execution time with \myCompilerNameSpace for most of the system sizes but remains orders of magnitude slower than \myCompilerName. % and exhibits significant solution errors particularly at the smallest size.

  \begin{figure*}[t]
    \centering
 %   \vspace{3pt}
    \includegraphics[width=1\textwidth]{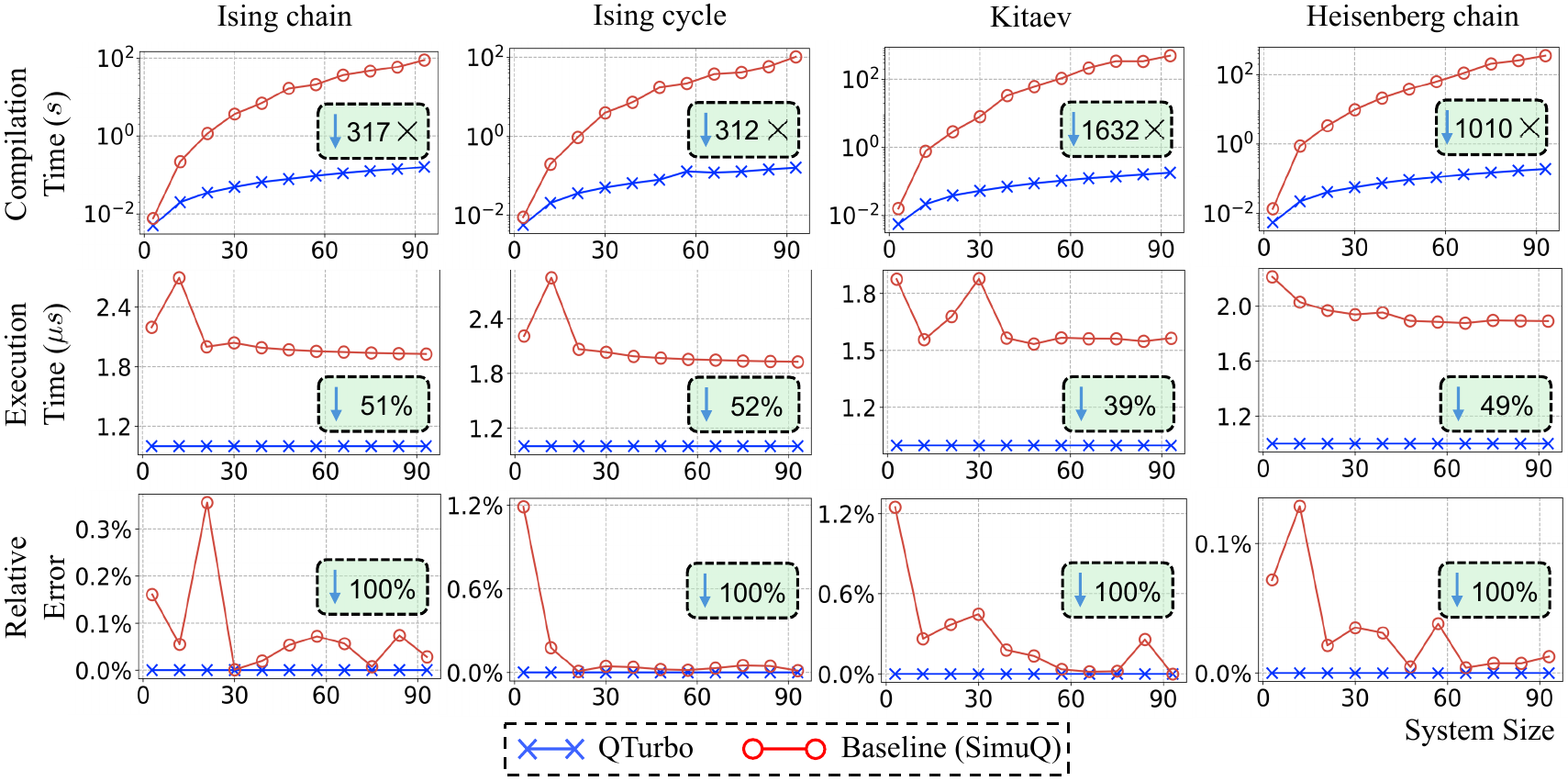}
   % \vspace{-13pt}
    \caption{Compilation result comparision for Heisenberg AAIS} 
   % \vspace{-12pt}
    \label{fig:CRH}
    \Description{}
\end{figure*}

Similarly, Figure~\ref{fig:CRH} shows the results for compiling the target models to the Heisenberg device. Our method demonstrates an average compilation speedup of $800 \times$, along with a $48\%$ reduction in execution time and a $100\%$ decrease in compilation error. The detailed breakdown can also be seen in each subgraph.

 % The overall results clearly demonstrate that
  In summary, \myCompilerNameSpace significantly reduces compilation time and scales effectively to large quantum systems. Additionally, the pulses generated by our compiler are highly optimized, exhibiting considerably shorter durations that enhance their robustness against noise.

  \subsection{Mapping and Time-Dependent Hamiltonian} \label{Sec:MappingandTimeDependentHamiltonian}

  In the preceding section, we focused on evaluating the optimization of the equation system—a fundamental component of the analog quantum simulation compiler. In this section, we discuss the challenges posed by the target-simulator mapping problem and the incorporation of a time-dependent Hamiltonian~\cite{peng2024simuq}. We examine how the observed accelerations/reductions resulting from optimizing the equation system contribute to overall performance improvements when factoring in these additional complexities.

  \begin{figure}[t]
    \centering
 %   \vspace{3pt}
    \includegraphics[width=1\columnwidth]{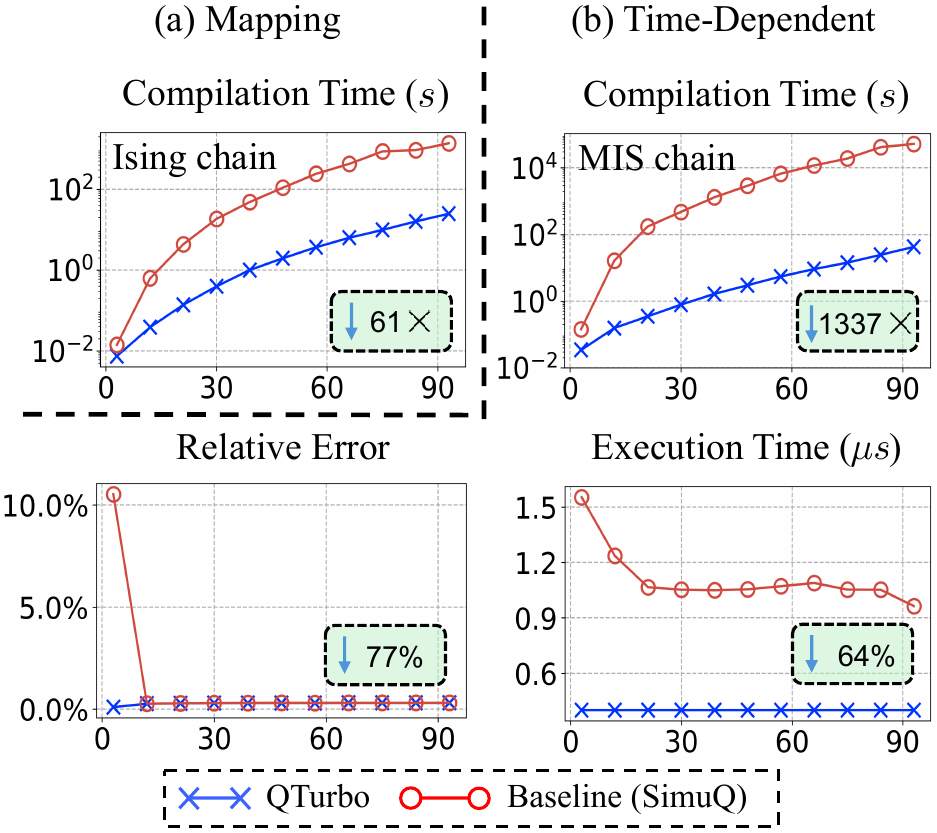}
   % \vspace{-13pt}
    \caption{Case study on mapping and time-dependent Hamiltonian} 
    \vspace{-12pt}
    \label{fig:M-T}
    \Description{}
\end{figure}

In practice, the physics models (see Table~\ref{tab:benchmarks}) have highly regular qubit coupling structures, such as a chain, a lattice, or a cycle, etc., and the mapping of such systems is usually not the bottleneck, compared with the equation solving.
   In Figure~\ref{fig:M-T} (a), an Ising chain model is compiled into a Rydberg device with an initially unknown mapping. %\jz{We employ a search with pruning approach, identical to that used by the SimuQ compiler, to determine the layout of the Ising model in the simulator.}
   We simply adopt the mapping in the baseline SimuQ~\cite{peng2024simuq} to determine the atom layout in the analog simulator.
   The result indicates that our method achieves $61 \times$ speedup in compilation, which is comparable to the results in Figure~\ref{fig:CRR}. The execution time and relative error results are the same as those in Figure~\ref{fig:CRR}.
   
   % and result for execution time and compilation error by $72\%$ and $2.6\%$, respectively. Notice that this result is closely related to the compilation result of Ising chain model in Figure~\ref{fig:CRR}.
   % We take extra compilation time here for finding the right target-to-simulator mapping.
   
   In Figure~\ref{fig:M-T} (b), a time-dependent Hamiltonian of MIS chain is also compiled into a Rydberg device. We divide the time interval into four segments and approximate the target Hamiltonian in each segment using a time-independent Hamiltonian. The results indicate that our method achieves an $1300 \times$ speedup in compilation while reducing both execution time and compilation error by $64\%$ and $77\%$, respectively.

  \subsection{ Real Device Experiments} \label{Sec:RealDeviceExecution}
  In this section we compare \myCompilerNameSpace and SimuQ on a real analog quantum simulator, QuEra's Aquila device. Our objective is to examine how reducing pulse length influences the accuracy of the quantum program.
%In this section, we validate the high performance of \myCompilerNameSpace by executing the pulse sequences on quantum hardware. 

Note that in real system studies, our metrics need to be practical observables that can be measured from the system. For Ising models, the average of the expectation of $Z$ observable across all qubits and the average of the expectation of $ZZ$ observable across all adjacent qubit pairs are widely used in physics study:
\begin{equation*}
   \mathcal{Z_{\text{avg}}} = \frac{1}{N} \sum_{i=1}^{N} \langle Z_{i}\rangle, \enspace \mathcal{ZZ_{\text{avg}}} = \frac{1}{N} \sum_{i=1}^{N} \langle Z_{i}Z_{i+1}\rangle.
\end{equation*} 
We measure these two values in our real system study.

\textbf{Firstly}, we choose a 12-atom Ising cycle model as the benchmark.
% \begin{equation*}
%     H = J \cdot \sum_{i=1}^{12} Z_{i}Z_{i+1} + h \cdot  \sum_{i=1}^{12} X_{i}.
% \end{equation*}
We set $J=0.157 \hspace{2pt} rad/\mu s$, and $h=0.785 \hspace{2pt} rad/\mu s$, with the target evolution time varying between $0.5 \hspace{2pt} \mu s $ and $1.0 \hspace{2pt} \mu s $. Furthermore, the maximum amplitude of $\Omega$ on the Aquila platform is limited to $6.28 \hspace{2pt} rad/ \mu s$. After compiling this analog program, \myCompilerNameSpace produces a pulse that achieves a $1.0 \hspace{2pt} \mu s$ target evolution while executing for only $0.25 \hspace{2pt} \mu s$ on a real device, whereas the SimuQ compiler requires $1.2 \hspace{2pt} \mu s$. The reduction is approximately $80\%$.

\begin{figure}[t]
    \centering
 %   \vspace{3pt}
    \includegraphics[width=1\columnwidth]{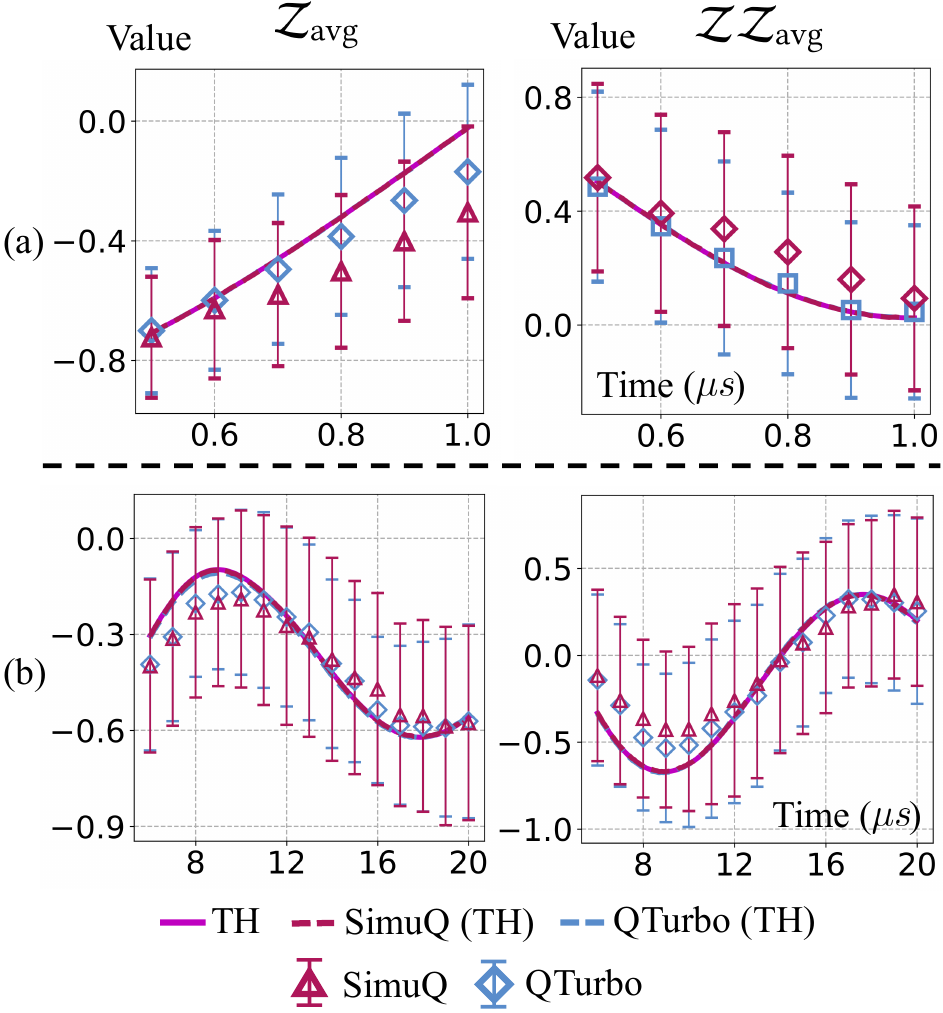}
   % \vspace{-13pt}
    \caption{Results on Aquila quantum computer} 
    \vspace{-12pt}
    \label{fig:RealExperiments}
    \Description{}
\end{figure}

Figure~\ref{fig:RealExperiments} (a) presents the values of metrics $\mathcal{Z_{\text{avg}}}$ and $\mathcal{ZZ_{\text{avg}}}$ with different target evolution time periods.
The theoretical values, labeled ``TH'' are computed using the QuTiP software~\cite{johansson2012qutip}. Additionally, we generate theoretical simulation values of QTurbo and SimuQ pulses using the Bloqade software~\cite{bloqade}, labeled as ``QTurbo (TH)'' and ``SimuQ (TH),'' respectively. We observe that in the absence of noise, both simulations align closely with the theoretical predictions, as the three theoretical curves are mostly overlapped. The real device execution results (each data point is averaged over 1000 shots) are labeled as ``QTurbo'' and ``SimuQ,'' respectively. 
It can be observed that the results from \myCompilerNameSpace compiler are much closer to the theoritical prediction compared with the results from SimuQ.
%As we can see, when various errors such as atomic random movement, pulse uncertainty, and other factors come into play, the execution time becomes a crucial factor. 
On average, our optimized pulse has achieved a $59\%$ (up to $81\%$) error reduction for metric $\mathcal{Z_{\text{avg}}}$, and $80\%$ (up to $94\%$) error reduction for metric $\mathcal{ZZ_{\text{avg}}}$.

\textbf{Secondly}, we choose a 6-atom PXP model as the benchmark.
% \begin{equation*}
%     H = J \cdot \Sigma n_{i}n_{i+1} + h \cdot \Sigma X_{i} = h \cdot \Sigma P_{i-1} X_{i} P_{i+1}.
% \end{equation*}
% \begin{figure}[t]
%     \centering
%  %   \vspace{3pt}
%     \includegraphics[width=1\columnwidth]{fig/EXP2.pdf}
%    % \vspace{-13pt}
%     \caption{Evolution Beyond Machine Time} 
%     \vspace{-12pt}
%     \label{fig:EvolutionBeyondMachineTime}
%     \Description{}
% \end{figure}
In this experiment, we set $J=1.26 \enspace rad/\mu s$, and $h=0.126 \enspace rad/\mu s$, with the target evolution time varying between $5 \mu s $ and $20 \mu s $. This specific ratio of  $J$ and $h$ guarantees that the Rydberg blockade is maintained thus validating the sufficiency of the PXP approximation. The maximum amplitude of $\Omega$ on the Aquila platform is limited to $13.8 \enspace rad/ \mu s$. With these constraints, a  $20 \mu s $ evolution can be compressed into just $0.4 \mu s$, while SimuQ use $3.4 \mu s$. The reduction is approximately $90\%$.

Figure~\ref{fig:RealExperiments} (b) presents the results. We observe $30\%$ (up to $87\%$) error reduction for metric $\mathcal{Z_{\text{avg}}}$, and $36\%$ (up to $92\%$) error reduction for metric $\mathcal{ZZ_{\text{avg}}}$. Additionally, it is important to note that the maximum execution time on Aquila is $4 \hspace{2pt} \mu s$, yet we compile for a target evolution time that significantly exceeds this limit—a key advantage of using an analog compiler.

\iffalse
\begin{figure}[t]
   \centering
   \vspace{3pt}
   \includegraphics[width=1\columnwidth]{fig/T-H.pdf}
   \vspace{-13pt}
   \caption{Evolution Time vs Model Parameter} 
   \vspace{-12pt}
   \label{fig:TH}
   \Description{}
\end{figure}

Furthermore, we can theoretically investigate the maximum evolution time that can be compressed within the machine's limitations, which ultimately depends on the model coefficients. Using the same PXP model, we generate a plot depicting the relationship between the maximum target evolution time and the model coefficient $h$. As shown in Figure~\ref{fig:TH}, we observe that a smaller interaction strength in the physical model allows for a longer compiled evolution time on the existing machine.

\fi
\section{Conclusion} \label{sec:Conclusion}

In summary, \myCompilerNameSpace effectively addresses existing inefficiencies in analog quantum compilation by strategically decomposing intricate global mixed systems into a global linear system combined with simpler, localized subsystems. Extensive experimental evaluations demonstrate notable advancements, including substantial improvements in compilation speed, execution efficiency, and the overall accuracy of quantum programs. Furthermore, rigorous tests conducted on real quantum hardware confirm \myCompilerName's robustness, illustrating significant enhancements in both fidelity and resilience against hardware noise and imperfections.

\section*{Acknowledgements}
This work was supported in part by the U.S. Department of Energy, Office of Science, Office of Advanced Scientific Computing Research under Contract No. DE-AC05-00OR22725 through the Accelerated Research in Quantum Computing Program MACH-Q project, the U.S. National Science Foundation CAREER Award No. CCF-2338773, and ExpandQISE Award No. OSI 2427020. GL was also supported by the Intel Rising Star Award. ER was supported by the U.S. Department of Energy (DOE) under Contract No. DE-AC02-05CH11231, through the National Energy Research Scientific Computing Center (NERSC), an Office of Science User Facility located at Lawrence Berkeley National Laboratory. EK and AM were supported by the U.S. Department of Energy (DOE) under Contract No. DE-AC02-05CH11231, through the Office of Advanced Scientific Computing Research Accelerated Research for Quantum Computing Program.

% use the ACM bibliography style
\bibliographystyle{ACM-Reference-Format}
\bibliography{refs}

\clearpage
\appendix
\section{Proof of Theorem~\ref{Th:errorbound}} \label{sec:appendix}

\textbf{[Error Bound of Compilation Process]} Suppose the $L_1$ error in solving the global linear equation system is bounded by $\epsilon_1$, and the $L_1$ error in solving each localized mixed equation system is bounded by $\epsilon_2^i$, then the total error in Equation (\ref{eq:errormetric}) is bounded by:
\begin{equation}
    \vert \vert M \vert \vert_{1} \cdot \sum_{i=1}^{K}\epsilon_2^i + \epsilon_1,
    \label{eq:errorbound}
\end{equation}
where $M$ is the matrix that governing the global linear system and $K$ is the number of localized mixed equation systems.

\begin{proof}
We define $\alpha$ and $x$ as the synthesized variables and amplitude variables, respectively, both represented as vectors. Notably, $\alpha$ depends on $x$ and can be expressed as $\alpha(x)$. 
Let $x^*$ denotes the solution we obtained, then 
 the Equation (\ref{eq:errormetric}) can be quantified as:
\begin{equation*}
    ||M\cdot \alpha(x^{*})-B_{tar}||_{1}
\end{equation*}
This quantity can be further decomposed as follows:
\begin{align*}
&\vert \vert M \cdot \alpha(x^{*}) - B_{tar} \vert \vert_{1} \\
     \leq &\vert \vert M \cdot \alpha(x^{*}) - M \cdot \alpha^* \vert \vert_{1} + \vert \vert M \cdot \alpha^* - B_{tar} \vert \vert_{1} \\
    \leq &\vert \vert M \vert \vert_{1} \cdot \vert \vert \alpha(x^{*}) - \alpha^* \vert \vert_{1} + \vert \vert M \cdot \alpha^* - B_{tar} \vert \vert_{1} \\
    \leq &\vert \vert M \vert \vert_{1} \cdot \left( \sum_{i=1}^{K} \vert \vert \alpha_i(x_i^{*}) - \alpha^*_i \vert \vert_{1} \right) + \vert \vert M \cdot \alpha^* - B_{tar} \vert \vert_{1}.
\end{align*}
In the second step, $\alpha^*$ represents the approximated solution to the global linear system, and the term $\vert \vert M \cdot \alpha^* - B_{tar} \vert \vert_{1}$ corresponds to the error introduced by the linear system itself. The term $\vert \vert \alpha(x^{*}) - \alpha^* \vert \vert_{1}$ represents the error introduced by the mixed equation system, which can be further decomposed into $K$ localized parts according to the dependence of the amplitude variables $x^*$. This separability allows the error to be expressed as the sum of individual error for the localized mixed equations, $\vert \vert \alpha_i(x_i^{*}) - \alpha_i^* \vert \vert_{1}$, as shown in the final step. By inserting the upper bound of each element:
\begin{equation*}
    \vert \vert \alpha_i(x_i^{*}) - \alpha^*_i \vert \vert_{1} \leq \epsilon_2^i, \enspace \vert \vert M \cdot \alpha^* - B_{tar} \vert \vert_{1} \leq \epsilon_1,
\end{equation*}
yields the result.

\end{proof}

\end{document}